\documentclass[runningheads]{llncs}
\usepackage[T1]{fontenc}
\usepackage{url}         
\usepackage{booktabs}    
\usepackage{amsfonts}      
\usepackage{nicefrac}       
\usepackage{amsmath,amssymb}
\usepackage{graphicx,wrapfig}
\usepackage{subfig}
\usepackage{tikz}
\usepackage{xcolor}
\usepackage{comment}
\usepackage{multirow}

\graphicspath{{Figs/}}

\begin{document}
\title{NPB-REC: Non-parametric Assessment of Uncertainty in Deep-learning-based MRI Reconstruction from Undersampled Data\thanks{To appear in the proceedings of the International Workshop on Machine Learning for Medical Image Reconstruction (MLMIR 2022) \protect\url{https://sites.google.com/view/mlmir2022}}}
\titlerunning{Assessment of Uncertainty in Deep-learning-based MRI Reconstruction}
%
\author{Samah Khawaled\inst{1}\orcidID{0000-0002-3053-0876} \and
Moti Freiman \inst{2}\orcidID{0000-0003-1083-1548} }

\authorrunning{S. Khawaled et al.}
%
\institute{Department of Applied Mathematics, Technion – Israel Institute of Technology \and
Faculty of Biomedical Engineering, Technion – Israel Institute of Technology \\
\email{ssamahkh@campus.technion.ac.il}, 
\email{moti.freiman@technion.ac.il}}
\maketitle              
\begin{abstract}
Uncertainty quantification in deep-learning (DL) based image reconstruction models is critical for reliable clinical decision making based on the reconstructed images.
We introduce ``NPB-REC'', a non-parametric fully Bayesian framework for uncertainty assessment in MRI reconstruction from undersampled ``k-space'' data. We use Stochastic gradient Langevin dynamics (SGLD) during the training phase to characterize the posterior distribution of the network weights. 
We demonstrated the added-value of our approach on the multi-coil brain MRI dataset, from the fastmri challenge, in comparison to the baseline E2E-VarNet with and without inference-time dropout. Our experiments show that  NPB-REC  outperforms the baseline by means of reconstruction accuracy (PSNR and SSIM of $34.55$, $0.908$ vs. $33.08$, $0.897$, $p<0.01$) in high acceleration rates ($R=8$). This is also measured in regions of clinical annotations. More significantly, it provides a more accurate estimate of the uncertainty that correlates with the reconstruction error, compared to the  Monte-Carlo inference time Dropout method (Pearson correlation coefficient of $R=0.94$ vs. $R=0.91$).
The proposed approach has the potential to facilitate safe utilization of DL based methods for MRI reconstruction from undersampled data. Code and trained models are available in \url{https://github.com/samahkh/NPB-REC}.


\end{abstract}

\section{Introduction}
Magnetic resonance imaging (MRI) is a noninvasive modality, which provides multi-planar images in-vivo through its sensitivity to the inherent magnetic properties of human tissue \cite{morris2018magnetic}. Although  MRI is the modality of choice in many clinical applications due to its excellent sensitivity to soft tissue contrast, its non-invasiveness, and the lack of harmful ionizing radiation, long acquisition times are a major limiting factor to achieve high spatial and temporal resolutions, reduce motion artifacts, improve the patient experience and reduce costs~\cite{fastmri}.

Reducing MRI acquisition time by under-sampling the ``k-space'' (i.e. Fourier domain) constitutes a key necessity in enabling advanced MRI applications such as cardiac and fetal imaging. Further, acceleration of MRI will also reduce MRI vulnerability to patient motion during the acquisition process. However, the under-sampled data results in aliasing artefacts in the reconstructed images. Early approaches rely upon Parallel Imaging (PI) \cite{griswold2002generalized,pruessmann1999sense} to reduce the acquisition time. This is done by utilizing multiple
receiver coils simultaneously to acquire multiple views and then combining them to construct the image. Other approaches reduce the acquisition time by sampling only a subset of measurements, i.e. under-sampling, and use a non-linear compressed sensing (CS) approach to reconstruct the MRI image from the under-sampled data \cite{candes2006compressive,lustig2007sparse}.

In the past few years, Deep-neural-networks (DNN) based models overcame classical reconstruction approaches in their ability to reconstruct high-quality MRI images from highly under-sampled data (i.e. $25\%$ or less) \cite{shaul2020subsampled,edupuganti2020uncertainty,eo2018kiki,akccakaya2019scan,tezcan2018mr,morris2018magnetic}. The Variational Network (VarNet) \cite{hammernik2018learning} solves an optimization problem by cascaded design of multiple layers, each solves a single gradient update step. The End-to-End Variational Network (E2E-VarNet) approach \cite{sriram2020end}, extends the VarNet model by estimation of the sensitivity maps within the network, which, in turn improves the quality of the reconstruction significantly at higher accelerations.
These models, however, 
do not provide posterior sampling neither enable uncertainty quantification, which are critical for clinical decision making based on the predicted images \cite{shaul2020subsampled}. 

Bayesian methods, such as Variational autoencoders (VAEs) and Monte Carlo dropout, are able to provide probabilistic interpretability and uncertainty quantification in MRI reconstruction \cite{edupuganti2020uncertainty,avci2021quantifying}. The VAE approach, however, is limited to specific DNN architectures. Further, it assumes a parametric distribution of the latent space in the form of a Gaussian distribution. 
 
In this work, we propose a non-parametric Bayesian DNN-based approach \cite{khawaled2022npbdreg,khawaled2020unsupervised} for MRI image reconstruction from under-sampled k-space data.  Our method is able to provide quantitative measures of uncertainty of the prediction by fully characterizing the entire posterior distribution of the reconstructed high-quality MRI images.
We achieve this by adopting the strategy of Stochastic gradient Langevin dynamics (SGLD) \cite{welling2011bayesian} to sample from the posterior distribution of the network weights \cite{cheng2019bayesian}. Specifically, we enable sampling by injecting Gaussian noise to the loss gradients during the training of the model. We save models with weights obtained after the \textit{``burn-in''} iteration, in which the training loss curve exhibits only small variations around its steady-state.
Then, at inference time, we estimate the statistics of the reconstructed image by averaging predictions obtained by the model with the saved weights.
Our contributions are: (1) the proposed approach can provide quantitative measures of uncertainty correlated with risk of failure in the MRI image predictions, and (2) can improve overall image reconstruction accuracy. These hypotheses were tested by experiments that were conducted on the publicly available fastMRI dataset\footnote{\url{https://fastmri.org/}}, for accelerated MRI reconstruction with the E2E-VarNet model \cite{sriram2020end} as the system's backbone. It is important to note that specifically, our main backbone, the E2E-VarNet model, doesn't provide posterior sampling neither enable uncertainty quantification.

\section{Methods}
\subsection{MRI Reconstruction}
In multi-coil acquisition, the MRI scanner consists of multiple receiver coils, where each of them partially acquires measurements in the k-space (frequency domain) \cite{majumdar_2015}. Each coil modulates the k-space samples according to its spatial sensitivity map to the MRI signal:
\begin{equation}
k_i=\mathcal{F}\left(S_ix\right)+\epsilon_i \:\: \forall i\in[1,..,N_c]    
\end{equation}
where $\mathcal{F}$ is the Fourier transform, $S_i$ denotes the corresponding sensitivity map and $N_c$ is the overall number of coils.
To accelerate the acquisition time, k-space measurements are under-sampled by selecting only a set from the entire k-space data, $\tilde{k_i}=M \circ k_i$, where $M$ is the corresponding binary mask that encodes the under-sampling operator. Restoration of the MRI image from the under-sampled data by performing an inverse Fourier transform on the under-sampled data leads to aliasing artifacts.

\begin{figure*}[t]
\centering
\scalebox{0.8}{	
\centering{
		\pgfdeclarelayer{background}
		\pgfdeclarelayer{foreground}
		\pgfsetlayers{background,main,foreground}
		
		\tikzstyle{sensor}=[draw, fill=gray!30, text width=5em, 
		text centered, minimum height=3em]
		\tikzstyle{ann} = [above, text width=4em]
		\tikzstyle{naveqs} = [sensor, text width=5em, fill=green!10, 
		minimum height=8em, rounded corners]
		\tikzstyle{naveqs1} = [sensor, text width=7em, fill=orange!20, 
		minimum height=8em, rounded corners]
		
		\def\blockdist{2.3}
		\def\edgedist{3}
		
		\begin{tikzpicture}
		\node (naveq) at (0,0) [naveqs] {UNet};
		\node (naveq1) at (0.03,0.03) [naveqs] {Rec. Nets};
		\node (naveq2) at (0.13,0.13) [naveqs] {Rec. Nets};
		\node (naveq3) at (0.23,0.23) [naveqs] {Rec. Nets};
		\node (naveq4) at (0.33,0.33) [naveqs] {Rec. Nets};
		\node (naveq41) at (4.3,0.3) [naveqs1,rotate=0] {Calculate Statistics};

		\path (naveq.140)+(-\blockdist,-0.4) node (gyros) [sensor] {K-space data};
		\path (naveq.140)+(9.3,0.1) node (reg) [sensor] {Rec. Image};
		\path (naveq.140)+(9.3,-1) node (sigma) [sensor] {Std. Map};
		
		\path [draw, ->] (gyros) -- node [above] {$\{\tilde{k_i}\}_{i=1}^{N_c}$} 
		(naveq.west |- gyros) ;
		\path (naveq.south west)+(-0.6,-0.4) node (INS) {};
		\draw[-] (naveq4) -- node [above](\edgedist,0) {{$\left\{ \hat{x}_{t}\right\} _{t_{b}}^{N}$}} ( naveq41.west |- naveq4);
    	\path [draw, <-] (reg) -- node [above] {$\bar{x}$} 
		(naveq41.east |- reg) ;
		 \path [draw, <-] (sigma) -- node [above] {$\Sigma$} 
		(naveq41.east |- sigma) ;
		\end{tikzpicture}
        }   }	
	\caption{Block diagram of the proposed NPB-REC system. The result of the SGLD-based training is a set of models with the same backbone network but diffident weights. At inference, the set of under-sampled k-space data,$\{\tilde{k_i}\}_{i=1}^{N_c}$,pass through each one of the backbone models. These models predict a set of reconstructed images, $\left\{ \hat{x}_{t}\right\} _{t_{b}}^{N}$.
	We then calculate the averaged reconstructed image and the pixel-wise std., $\bar{x}$ and $\Sigma$, respectively. The average is used as the most probable reconstruction prediction and the $\Sigma$ is utilized for uncertainty assessment. }
	\label{fig:bddiagram}
\end{figure*}
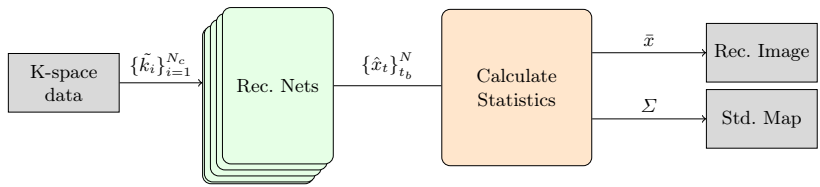

\subsection{Non-parametric Bayesian MRI Reconstruction}\label{subsec:offlinetraining}
 Our goal is to characterize the posterior distribution:
\begin{equation}
\hat{\theta} \sim {P\left(\theta|x,\hat{x},\left\{\tilde{k_i}\right\}_{i=1}^{N_c}\right) \propto P\left(x,\hat{x},\left\{\tilde{k_i}\right\}_{i=1}^{N_c}|\theta\right)P\left(\theta\right)} \label{eq:4}
\end{equation} 
where, $\hat{x}$, $x$, and $\{\tilde{k_i}\}_{i=1}^{N_c}$ are the reconstructed image, the ground truth (fully sampled) and the under-sampled k-space measurements, respectively. $\theta$ are the network weights that we optimize. 
 We treat the network weights as random variables and aim to sample the posterior distribution of the model prediction. To this end, we incorporate a noise scheduler that injects a time-dependent Gaussian noise to the gradients of the loss during the optimization process. At every training iteration, we add Gaussian noise to the loss gradients. Then, the weights are updated in the next iteration according to the "noisy" gradients. This noise schedule can be performed with any stochastic optimization algorithm during the training procedure. 
In this work we focused on the formulation of the method for the \texttt{Adam} optimizer. 

The reconstruction network predicts: $\hat{x}=f_\theta\left(\left\{\tilde{k_i}\right\}_{i=1}^{N_c}\right)$. 
We denote the loss gradients by:
\begin{equation}
g^t\overset{\triangle}{=}\nabla_{\theta}L^t\left(x,f_\theta\left(\left\{\tilde{k_i}\right\}_{i=1}^{N_c}\right)\right)
\end{equation}
where $L^{t}$ is the reconstruction loss at the training iteration (epoch) $t$. At each training iteration, a Gaussian noise is added to $g$:
\begin{equation}
\tilde{g}^{t}\leftarrow g^{t}+\textbf{N}^{t}
\end{equation}
where $\textbf{N}^{t}\sim\mathcal{N}(0,s^t)$, $s^t$ is a  user-selected parameter that controls the noise variance (can be time-decaying or a constant). We selected $s^t$ equal to the \texttt{Adam} learning rate.

Lastly, we save the weights of the network that were obtained in iterations $t\in \left[t_b,N\right]$, where $N$ is the overall number of iterations and $t_b$ is the SGLD-parameter, which should be larger than the cut-off point of the \textit{burn-in} phase. One should sample weights obtained in the last $t_b,..,N$ iterations, where the loss curve has converged. 

We exploit the network weights that were obtained after the \textit{burn-in} phase, i.e. in the last $t_b,..,N$ iterations, $\left\{ \theta\right\}_{t_{b}}^{N}$. Fig.~\ref{fig:bddiagram} illustrates the operation of the NPB-REC system at the inference phase.  
We sample a set of reconstructed images  $\left\{ \hat{x}\right\}_{t_{b}}^{N}$, obtained by feed-forwarding the under-sampled k-space data $\{\tilde{k_i}\}_{i=1}^{N_c}$ to the reconstitution models with the weights $\left\{ \theta\right\}_{t_{b}}^{N}$.
Then, when we have a new image to reconstruct, we estimate the averaged posterior image:
\begin{equation}
\bar{x}=\frac{\sum_{t=t_b}^{N} \hat{x}_t}{N-t_b} \label{eq:avg}
\end{equation} 
In addition, we quantify the std. of the reconstruction, which is used to characterize the uncertainty.

\subsection{The Reconstruction Network}
The backbone of our reconstruction system is based on the E2E-VarNet model \cite{sriram2020end}. E2E-Varnet contains multiple cascaded layers, each applies a refinement step in the k-space according to the following update: 
\begin{equation}
    k^{m+1} = k^m - \eta^t M \left(k^m-\tilde{k}\right)+G\left(k^m\right) \label{eq:kspacesolve}
\end{equation} 
where $k^m$ and $k^{m+1}$ are the input and output to the m-th layer, respectively. $G$ is the refinement module: $ G = \mathcal{F}\circ\mathcal{E}\circ \text{CNN}(\mathcal{R}\circ\mathcal{F}^{-1}(k^m)$.
 CNN is a DNN-network, which maps a complex image input to a complex output, $\mathcal{E}$ and $\mathcal{R}$ are the expand and reduce operators. $\mathcal{E}$ computes the corresponding image seen by each coil: $\mathcal{E}(x) = (S_{1}x,...,S_{N_c}x)$ and its inverse operator, $\mathcal{R}$, integrates the multiple coil images $\mathcal{R}(x_1,...,x_{N_c})=\sum S_ix_i$. 
Similarly to the design of E2E-VarNet \cite{sriram2020end}, a U-Net is used for the CNN \cite{ronneberger2015u}. The sensitivity maps $S_1,...,S_{N_c}$ are estimated as part of the reconstruction using a CNN that has the same architecture as the CNN in the cascades. After applying a cascaded layers to the k-space input, as described in \eqref{eq:kspacesolve}, we obtain the final reconstructed image, $\hat{x}$,  by root-sum-squares (RSS) reduction of the image-space representation of the final layer output: $\hat{x} = \sqrt{\sum_{i=1}^{N_c}|\mathcal{F}^{-1}k^{T}_i|^2}$.

\section{Experiments}
\subsection{Database}
In our experiments, we used multi-coil data of brain MRI images, adapted from the publicly available fastMRI database
  for training our system; we used the validation and training datasets for the brain multi-coil challenge. We excluded a subset from the validation set that contained clinical pathology annotations taken from fastMRI+ \cite{zhao2021fastmri} and used it in the evaluation of our method, where no overlap between subjects belong to these sets. This is due to the fact that the ground truth of the test set is not publicly available and it is interesting to demonstrate our NPB-REC method on region of interest (ROIs) and to quantify both the reconstruction accuracy and uncertainty in these ROIs. 

The training, validation, and inference sets include $4469$, $113$, and $247$ images of size $16\times 320\times 320$ and $20$ coils. From these datasets, we generate $71504$,  $1808$ 2D images and $2141$ slices with annotations for the training, validation and the evaluation phases, respectively.
The inputs of the network, $k^0$ in \eqref{eq:kspacesolve}, are under-sampled k-space inputs that were generated from the fully-sampled datasets with two types of masks: \textit{equispaced}  and \textit{random}. The former samples $l$ low-frequency
lines from the center of k-space and every R-th line from the higher frequencies, to make the acceleration rate equal to $R$. Whereas, the latter samples a fraction of the full width of k-space for the central k-space corresponding to low frequencies and selects uniformly at a subset of a higher frequency line such that the overall acceleration is $R$. 

\subsection{Experimental Setup}
To conduct a quantitative comparison, we trained three models: (1) E2E-Varnet \cite{sriram2020end} trained with the proposed NPB-REC method (section~\ref{subsec:offlinetraining}), (2) The baseline E2E-VarNet model \cite{sriram2020end}, and; (3) E2E-VarNet trained with Dropout of probability $0.001$ and Monte Carlo averaging used at inference. Higher values of Dropout probabilities led to instabilities in the network loss during training. The three aforementioned models have the same architecture as described in \cite{sriram2020end} with $T=8$ cascaded layers. Dropout layers were incorporated only to model (3), whereas the first two configurations where trained without adding it. For the three models, we used $1-SSIM$ as a training loss and \texttt{Adam} optimizer with learning rate set to $lr=0.001$. The total number of epochs (training iterations) was set to $40$ and the batch size equal to $1$. 
In our experiments, we selected  a standard deviation, $s^t=lr$ for the injected noise variance.
The network parameters are then updated according to the \texttt{Adam} update rule. In the training, we generated under-sampled inputs by multiplying with \textit{random} masks of acceleration rate $R=4$. At inference, we use both \textit{random} and \textit{equispaced} masks of acceleration rates $R=4$ and $R=8$, to evaluate the ability of the system to generalize.  

\begin{wrapfigure}{rt!}{5.5cm}
	\includegraphics[width=5cm]{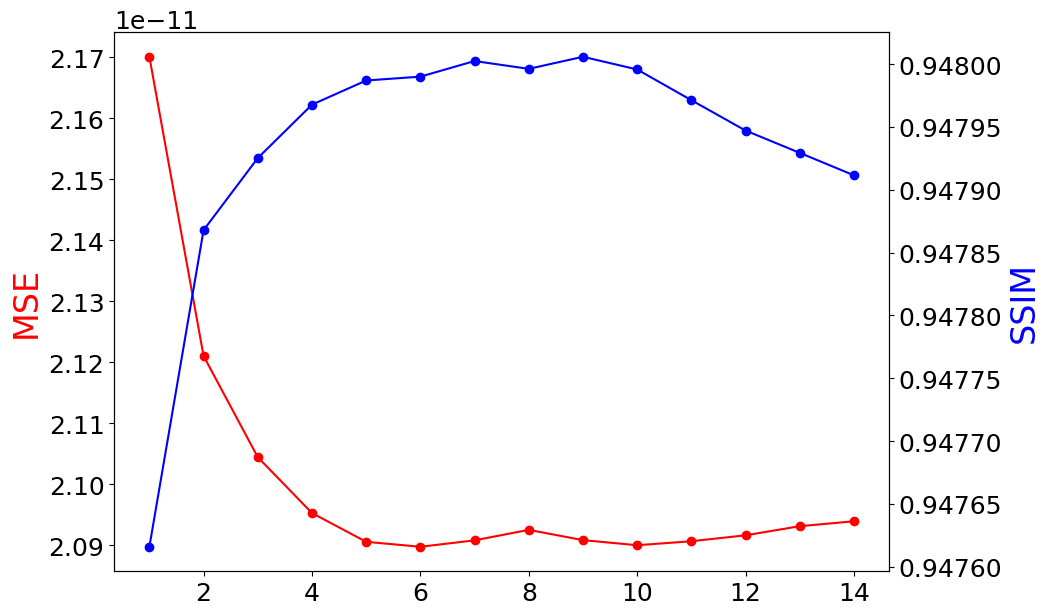}
	\caption{MSE and SSIM vs. $N-t_b$, the number of models used in the averaging, obtained on subset from the inference set (32 images sampled randomly).}
\label{fig:masks}
\end{wrapfigure} 

\textbf{Hyper-parameters Selection:} At inference, we sample a set of reconstructed images obtained by passing the under-sampled k-space inputs to the models with the weights $\left\{ \theta\right\}_{t_{b}}^{N}$, i.e. obtained in the last $N-t_b$ iterations. $t_b$ should be selected such that the training loss in the last $N-t_b$ is stable and has only slight variations around its steady state value. In our experiments, we performed a hyper-parameter tuning on $t_b$ and selected the last $N-t_b$ that obtained the best quantitative reconstruction performance. 
 
 Fig.~\ref{fig:masks} presents the MSE and SSIM metrics for a range of $N-t_b$ that varies from 1 to 10. Although this range of $N-t_b$ values show similar MSE and SSIM metrics, we selected $N-t_b=9$. This is due to the fact that it shows a slight improvement and with having $9$ samples we can calculate robust statistics. 

The final reconstructed image is calculated by averaging these $9$ samples, predicted from our model, as mentioned in \eqref{eq:avg}. Additionally, we estimate a 2D uncertainty map by calculation of the pixel-wise std. of these predictions. 
For fair comparison, we used the same number of predictions in Monte Carlo sampling at the inference phase, but with enabled Dropout layers. 
We assessed the accuracy of the registration models by calculating PSNR and SSIM between the  reconstructed and the ground truth, for all pairs of images in the test set.
 \begin{figure*}[t!]
\centering
	\begin{minipage}[b]{0.18\linewidth}
		\centering
		\includegraphics[width=2.1cm]{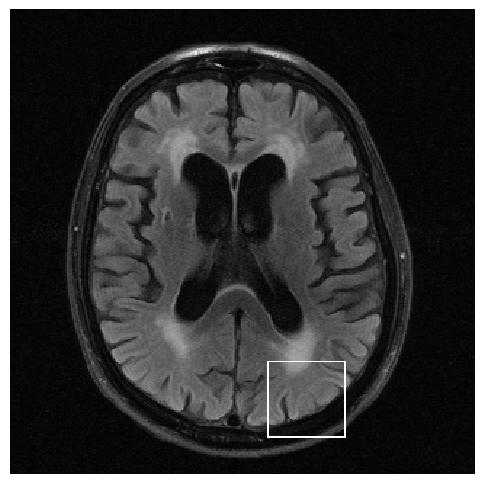}
	\end{minipage} 
	\begin{minipage}[b]{0.18\linewidth}
		\centering
		\includegraphics[width=2.1cm]{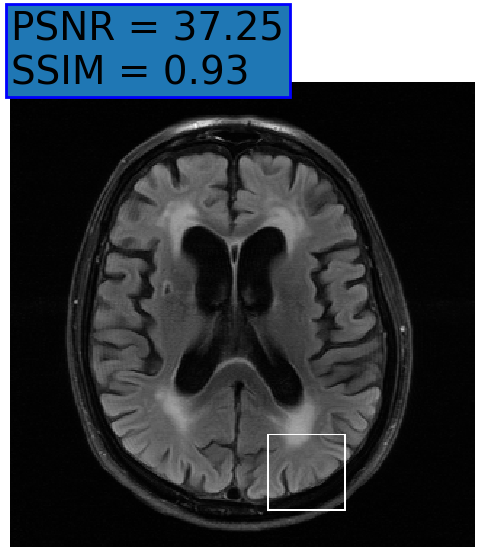}
	\end{minipage}
	\begin{minipage}[b]{0.18\linewidth}
		\centering
		\includegraphics[width=2.1cm]{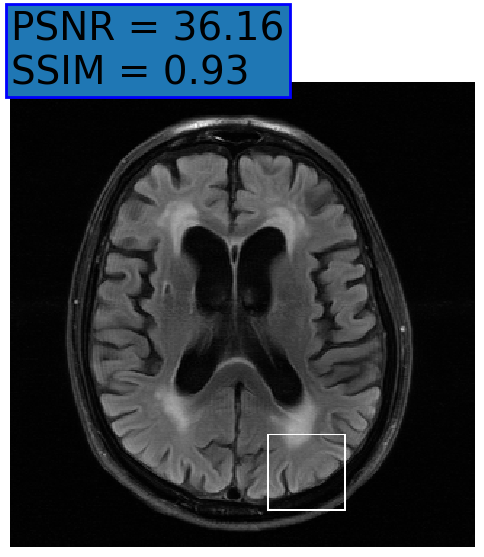}
	\end{minipage} 
	\begin{minipage}[b]{0.18\linewidth}
		\centering
		\includegraphics[width=2.1cm]{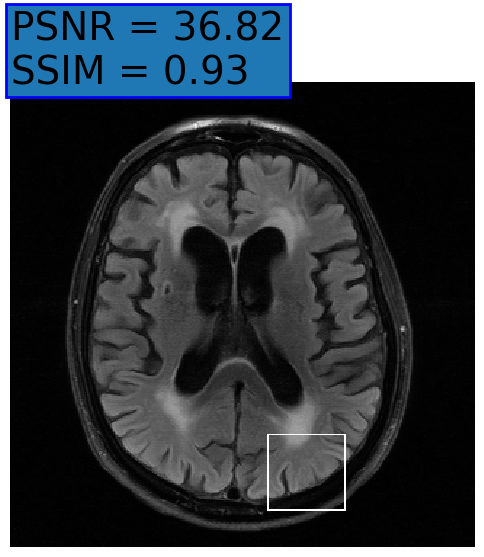}
	\end{minipage} 
		\begin{minipage}[b]{0.18\linewidth}
		\centering
		\includegraphics[width=2.1cm]{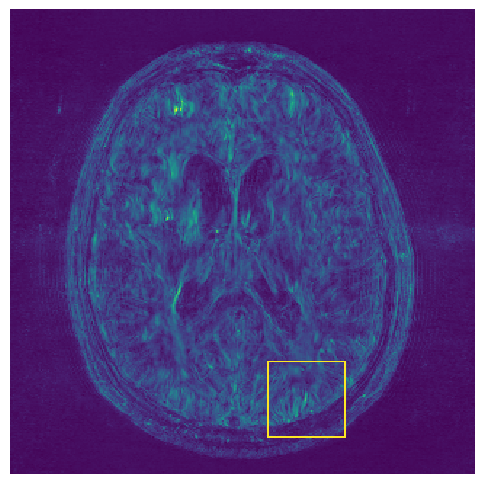}
	\end{minipage}
	\\
	\begin{minipage}[b]{0.18\linewidth}
		\centering
		\includegraphics[width=2.1cm]{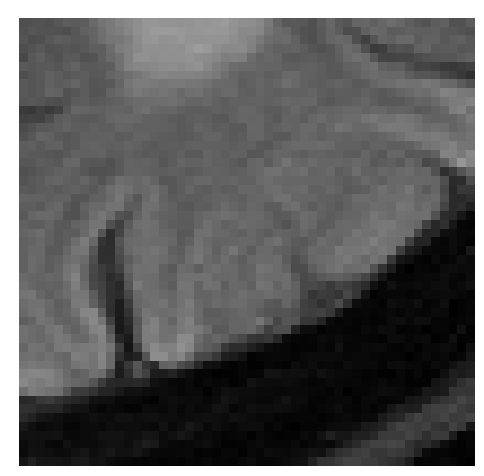}
	\end{minipage} 
	\begin{minipage}[b]{0.18\linewidth}
		\centering
		\includegraphics[width=2.1cm]{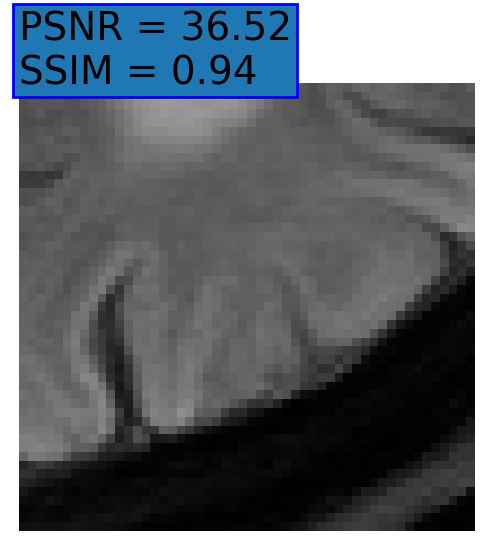}
	\end{minipage} 
		\begin{minipage}[b]{0.18\linewidth}
		\centering
		\includegraphics[width=2.1cm]{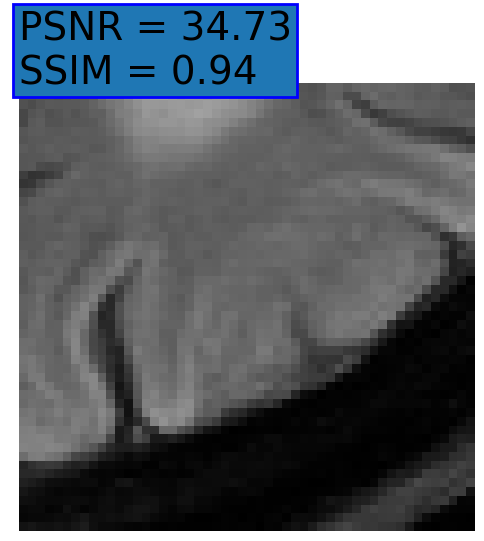}
	\end{minipage} 
		\begin{minipage}[b]{0.18\linewidth}
		\centering
		\includegraphics[width=2.1cm]{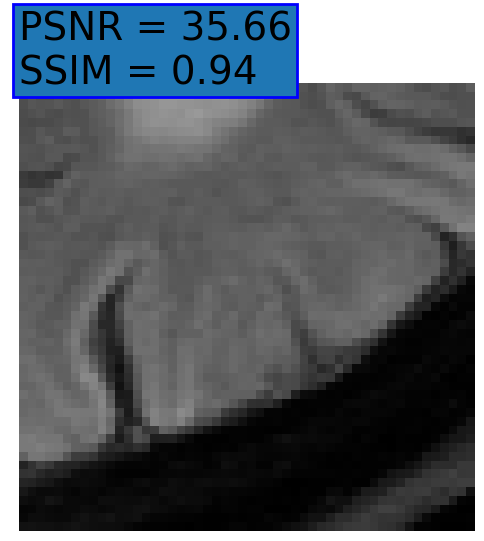}
	\end{minipage} 
		\begin{minipage}[b]{0.18\linewidth}
		\centering
		\includegraphics[width=2.1cm]{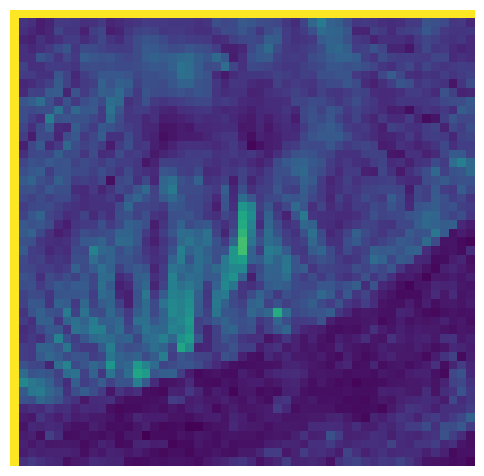}
	\end{minipage}
    \\
	\begin{minipage}[b]{0.18\linewidth}
		\centering
		\includegraphics[width=2.1cm]{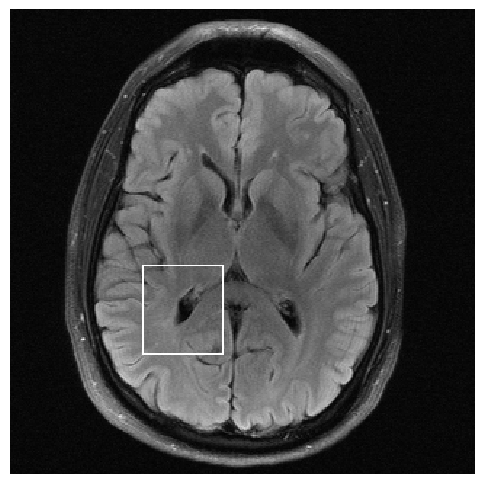}
	\end{minipage} 
	\begin{minipage}[b]{0.18\linewidth}
		\centering
		\includegraphics[width=2.1cm]{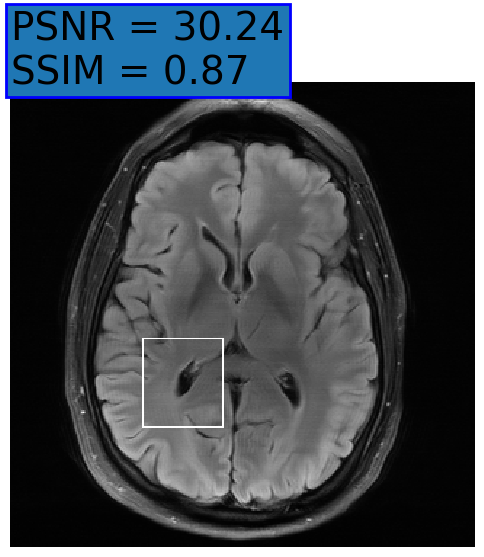}
	\end{minipage}
	\begin{minipage}[b]{0.18\linewidth}
		\centering
		\includegraphics[width=2.1cm]{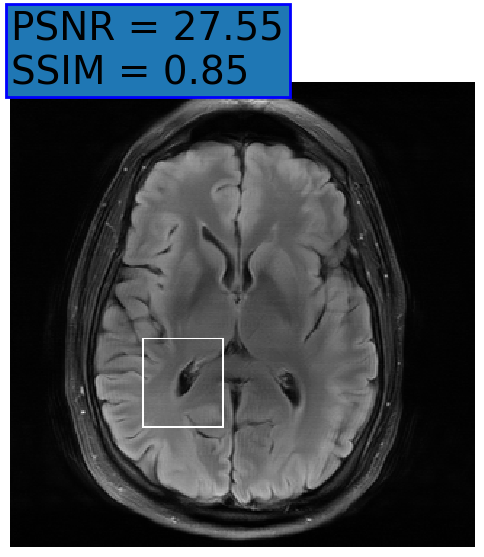}
	\end{minipage} 
	\begin{minipage}[b]{0.18\linewidth}
		\centering
		\includegraphics[width=2.1cm]{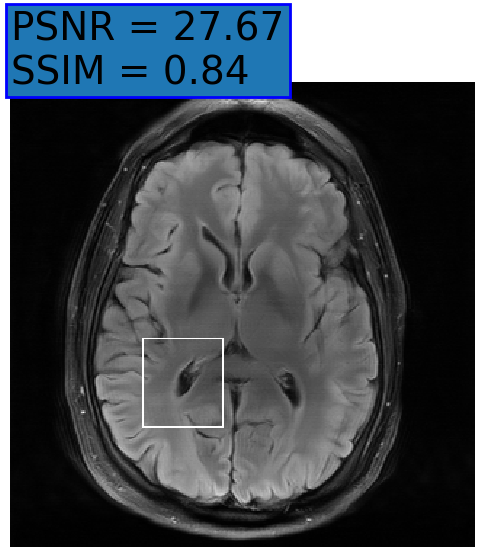}
	\end{minipage} 
		\begin{minipage}[b]{0.18\linewidth}
		\centering
		\includegraphics[width=2.1cm]{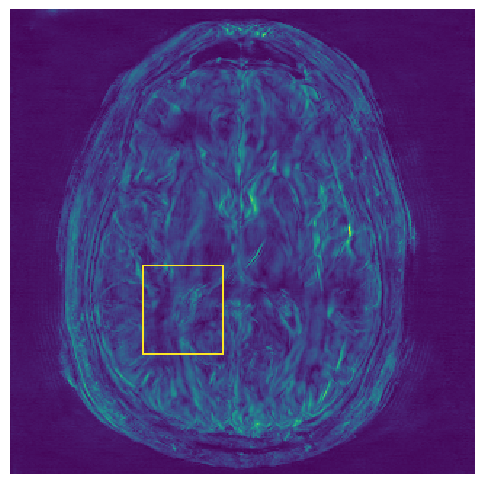}
	\end{minipage}
	\\
		\begin{minipage}[b]{0.18\linewidth}
		\centering
		\subfloat[\label{fig3:b}\footnotesize GT]{\includegraphics[width=2.1cm]{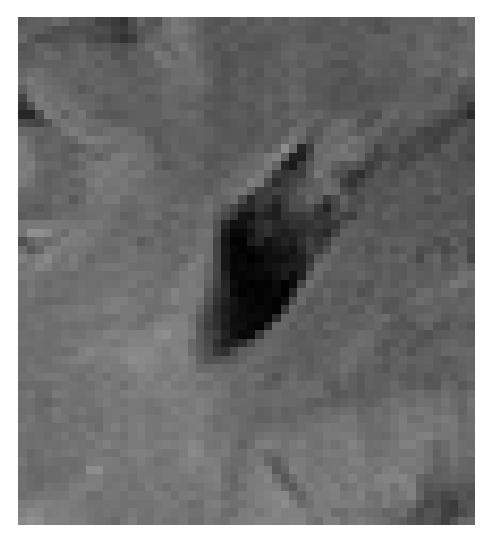}}
	\end{minipage} 
	\begin{minipage}[b]{0.18\linewidth}
		\centering
		\subfloat[\footnotesize NPB-REC]{\includegraphics[width=2.1cm]{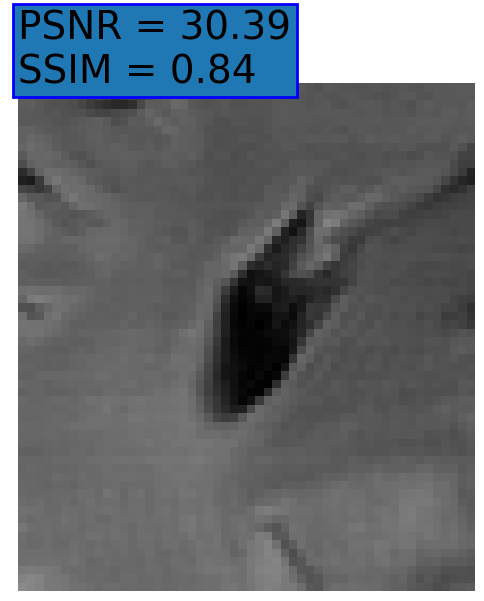}}
	\end{minipage} 
		\begin{minipage}[b]{0.18\linewidth}
		\centering
		\subfloat[\label{fig2:c}\footnotesize baseline]{\includegraphics[width=2.1cm]{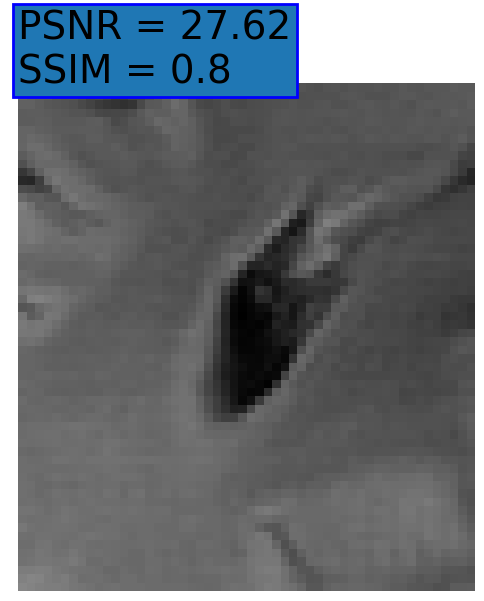}}
	\end{minipage} 
		\begin{minipage}[b]{0.18\linewidth}
		\centering
		\subfloat[\label{fig2:d}\footnotesize Dropout]{\includegraphics[width=2.1cm]{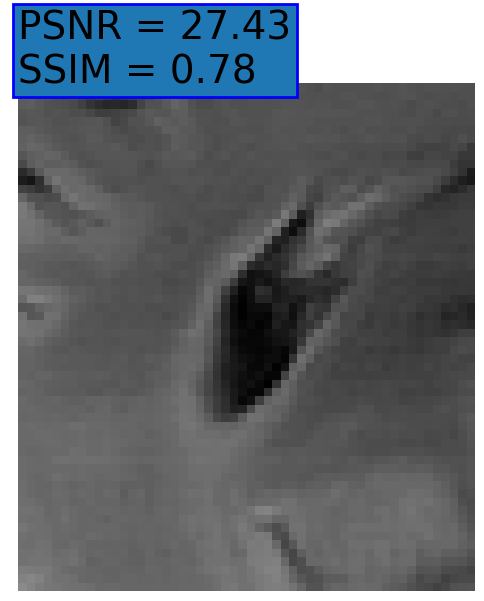}}
	\end{minipage} 
		\begin{minipage}[b]{0.18\linewidth}
		\centering
		\subfloat[\label{fig2:d}\footnotesize Std.]{\includegraphics[width=2cm]{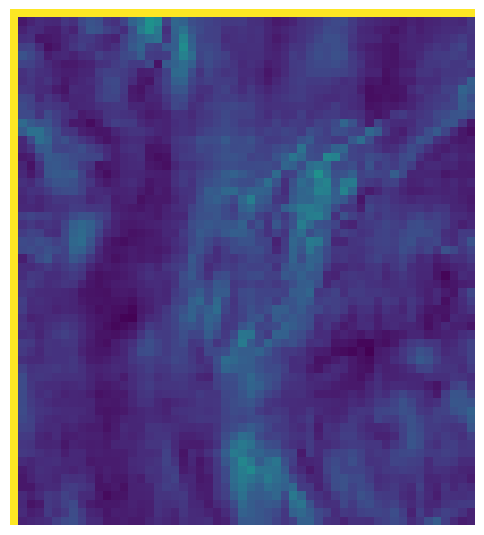}}
	\end{minipage}
	\caption{Examples of Reconstruction Results. Rows 1 and 3: The Ground truth (GT) fully sampled image, the reconstructed images obtained by the three models (1-3), NPB-REC, baseline, E2E-VarNet trained with Dropout, and the Std. map derived from our method for acceleration rates $R=4$, $R=8$, respectively. Rows 2 and 4: The corresponding annotated ROIS of the edema and resection cavity. }\label{fig:RecMRIexample}	
\end{figure*} 
 
\subsection{Results}
Fig.~\ref{fig:RecMRIexample} presents examples of reconstruction results obtained by (1) our NPB-REC approach, (2) the baseline, and (3) Monte Carlo Dropout, for equispaced masks with two different acceleration rates $R=4$ and $R=8$. 
 Table~\ref{tab:results} presents the mean PSNR and SSIM metrics, calculated over the whole inference set, for the three models. our NPB-REC approach achieved significant improvements over the other methods in terms of PSNR and SSIM (Wilcoxon signed-rank test, p$\ll$1e-4). The improvement in the reconstruction performance can be noted both quantitatively from the metrics especially for masks with acceleration rate $R=8$ and qualitatively via the images of annotations, where our results shows less smoothness than that obtained by Dropout.
 
\begin{figure*}[t!]
	\begin{minipage}[b]{0.32\linewidth}
		\centering
		\subfloat[\label{fig3:a}\footnotesize NPB-REC]{
		\includegraphics[height=2.1cm,width=3.3cm]{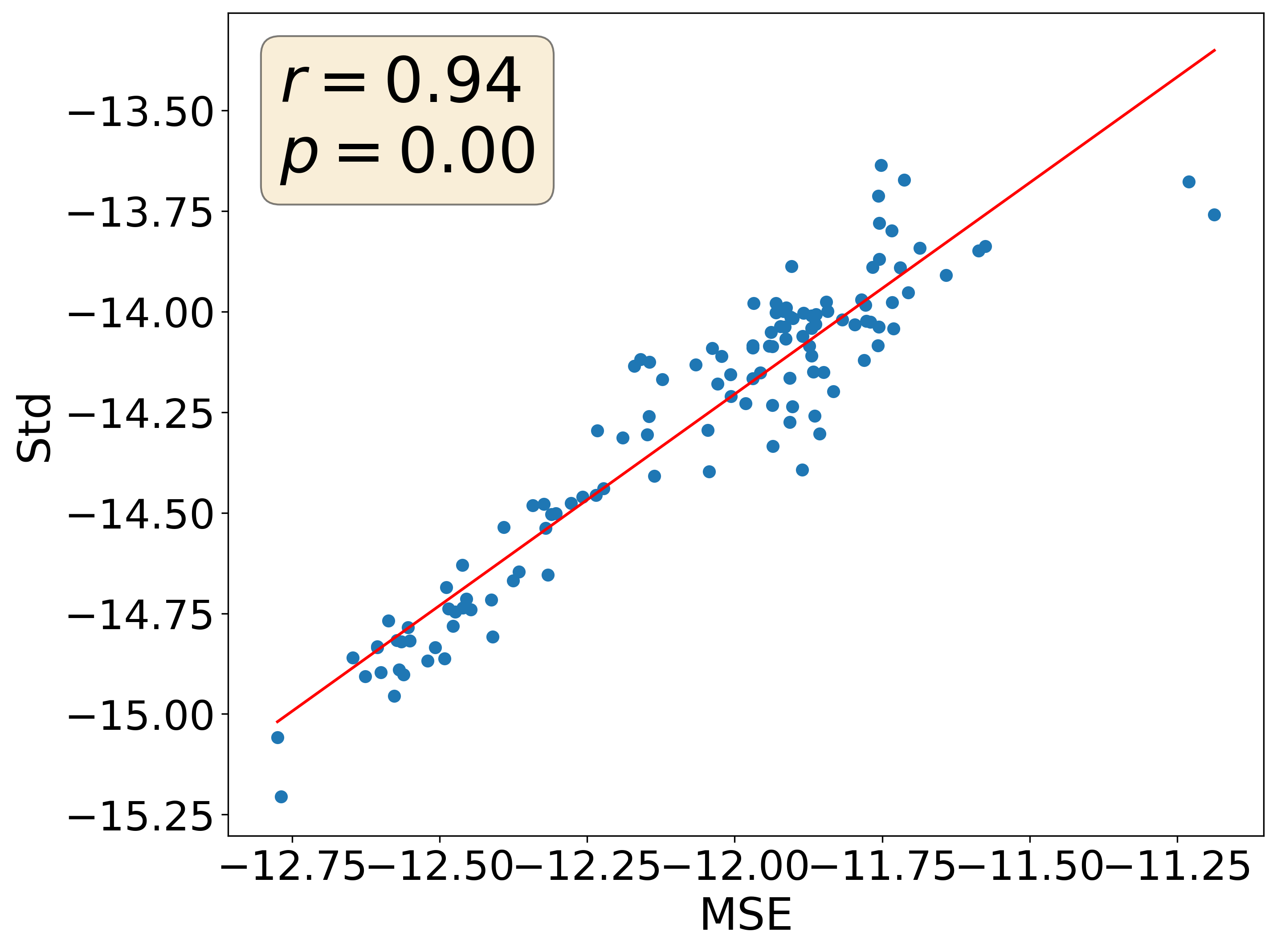}}
	\end{minipage}
	\begin{minipage}[b]{0.32\linewidth}
		\centering
		\subfloat[\label{fig3:b}\footnotesize Dropout]{
		\includegraphics[height=2.1cm,width=3.3cm]{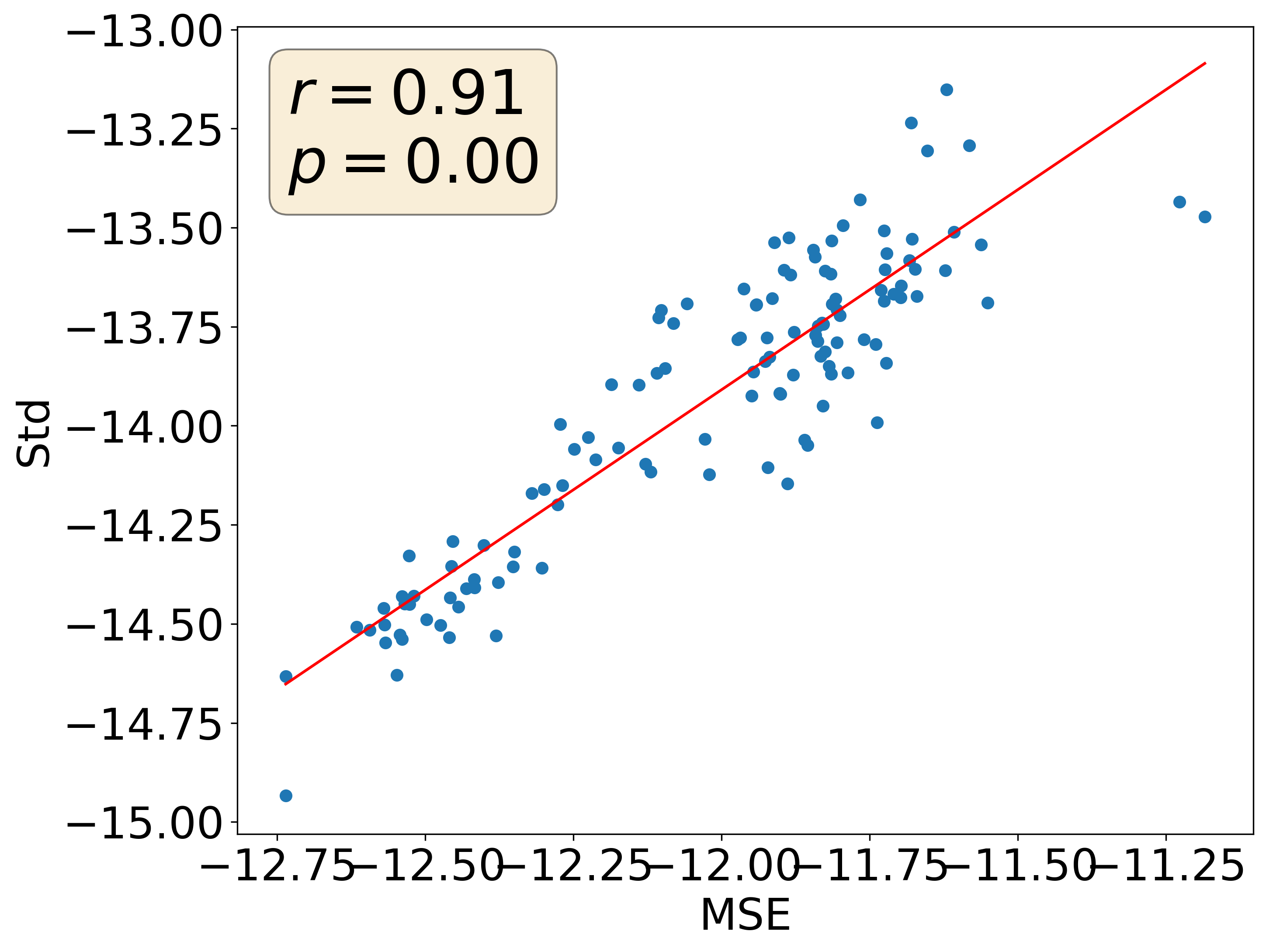}}
	\end{minipage}	
	\begin{minipage}[b]{0.32\linewidth}
		\centering
		\subfloat[\label{fig3:c}  Uncertainty vs. $R$]{
		\includegraphics[height=2.1cm,width=3.3cm]{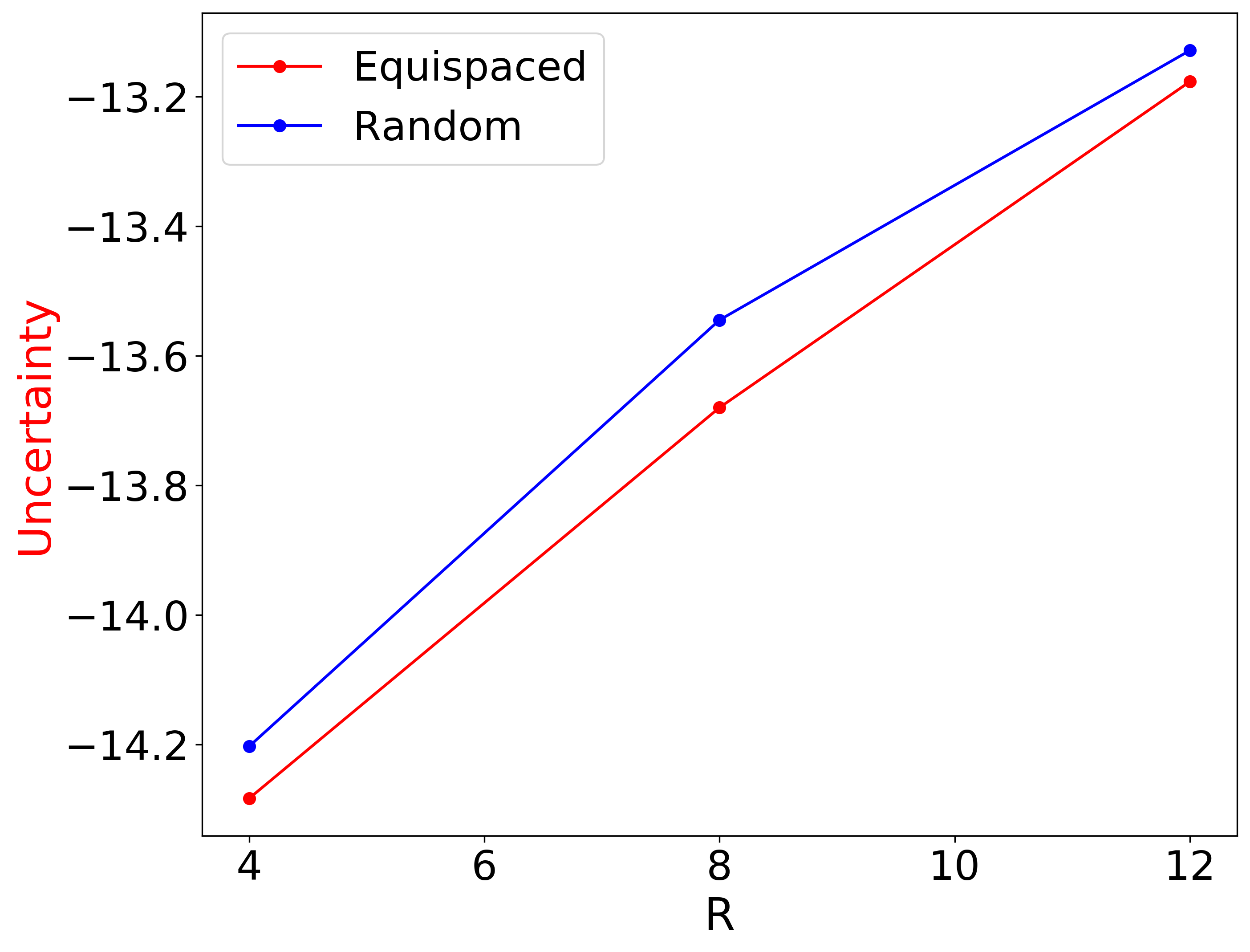}}
	\end{minipage}
	\caption{Uncertainty Assessment. Scatter plots of the mean value of Std. estimate versus the MSE metric, calculated between the reconstructed and the ground truth, in log scale, for our NPB-REC method \protect\subref{fig3:a} and Monte Carlo Dropout \protect\subref{fig3:b}. \protect\subref{fig3:c} Our measure of uncertainty versus the acceleration rate.  }\label{fig:unc}	
\end{figure*}

\textbf{Meaning of the uncertainty Measures:}
We calculated the mean value of the Std. maps, obtained by our method and Monte Carlo Dropout, for all images in the inference set and utilize it as uncertainty measure.
The correlation between these uncertainty measures and reconstruction error (MSE) are depicted in Fig.~\ref{fig:unc}. Compared to Dropout, our NPB-REC uncertainty measure exhibits higher correlation with the Reconstruction error (Pearson correlation coefficient of  $R=0.94$ vs. $R=0.91$). Further, Fig.~\ref{fig:unc}\protect\subref{fig3:c} demonstrates that higher acceleration rates  increases the uncertainty measure. 
These outcomes, in turn, indicate the ability of our uncertainty measure to detect unreliable reconstruction performance.

\begin{table}[t]
\caption{Reconstruction Accuracy. Rows top to bottom: PSNR and SSIM metrics calculated on the annotated anatomical ROIs (denoted by 'A') with mask of acceleration rate $R=4$, the whole physical images (denoted by 'W') with masks of acceleration rate $R=4$, $R=8$, respectively. 'r' and 'e' stands for \textit{random} and \textit{equispaced} mask types.} \label{tab:results}
\setlength{\tabcolsep}{9pt}
\centering
\resizebox{\textwidth}{!}{%
\begin{tabular}{@{}lcccccccc@{}}
\toprule
\footnotesize
&                                                        &                             & \multicolumn{2}{c}{\textbf{NPB-REC}}                                & \multicolumn{2}{c}{\textbf{Baseline}}                                                   & \multicolumn{2}{c}{\textbf{Dropout}}                                                    \\ \midrule
\multicolumn{1}{l}{}                    & \multicolumn{1}{c}{\textbf{R}}                                                                     & \multicolumn{1}{c}{\textbf{M}}   & \multicolumn{1}{c}{\textbf{PSNR}}            & \multicolumn{1}{c}{\textbf{SSIM}}               & \multicolumn{1}{c}{\textbf{PSNR}}            & \multicolumn{1}{c}{\textbf{SSIM}}               & \multicolumn{1}{c}{\textbf{PSNR}}            & \multicolumn{1}{c}{\textbf{SSIM}}               \\ \midrule
\multicolumn{1}{l}{\multirow{2}{*}{A}} & \multicolumn{1}{c}{\multirow{2}{*}{\begin{tabular}[c]{@{}c@{}}4\end{tabular}}} & \multicolumn{1}{c}{r} &  \multicolumn{1}{c}{$\mathbf{30.04\pm6.78}$} & $\mathbf{0.87\pm0.18}$ & \multicolumn{1}{c}{$29.91\pm6.87$} & $0.867\pm0.182$   & \multicolumn{1}{c}{$29.5\pm 6.844$}  & $0.858\pm 0.19$  \\ \cmidrule(l){3-9}  
\multicolumn{1}{l}{}                   & \multicolumn{1}{c}{}                                                                       & \multicolumn{1}{c}{e}   & \multicolumn{1}{c}{$\mathbf{32.22\pm 6.94}$} & $\mathbf{0.914\pm0.143}$  & \multicolumn{1}{c}{$32.02\pm 7.35$} & $0.911\pm0.143$  & \multicolumn{1}{c}{$31.57\pm6.89$} & $0.905\pm0.151$    \\ \midrule
\multicolumn{1}{l}{\multirow{2}{*}{W}} & \multicolumn{1}{c}{\multirow{2}{*}{4}}                      & \multicolumn{1}{c}{r} & \multicolumn{1}{c}{$\mathbf{40.24\pm 6.19}$} & $\mathbf{0.947\pm0.081}$ & \multicolumn{1}{c}{$40.17\pm 6.19$} & $\mathbf{0.947\pm0.081}$  & \multicolumn{1}{c}{$39.86\pm6.10$} & $0.945\pm0.082$ 
\\ \cmidrule(l){3-9} 
\multicolumn{1}{l}{}                   & \multicolumn{1}{c}{}                                                                       & \multicolumn{1}{c}{e}   &  \multicolumn{1}{c}{$41.61\pm6.28$} & $\mathbf{0.955\pm0.073}$ & \multicolumn{1}{c}{$\mathbf{41.64\pm6.28}$} & $\mathbf{0.955\pm0.074}$ & \multicolumn{1}{c}{$41.22\pm6.0$} & $0.953\pm0.074$ 
\\ \midrule
\multicolumn{1}{l}{\multirow{2}{*}{W}} & \multicolumn{1}{c}{\multirow{2}{*}{8}}                                                   & \multicolumn{1}{c}{r} & \multicolumn{1}{c}{$\mathbf{32.23\pm6.63}$} & $\mathbf{0.881\pm0.11}$ & \multicolumn{1}{c}{$31.21\pm6.2$} & $0.87\pm0.108$  & \multicolumn{1}{c}{$30.63\pm5.91$} & $0.865\pm0.11$  
\\ \cmidrule(l){3-9} 
\multicolumn{1}{l}{}                   & \multicolumn{1}{c}{}    & e  & \multicolumn{1}{c}{$\mathbf{34.55\pm5.01}$} & $\mathbf{0.908\pm0.09}$ & \multicolumn{1}{c}{$33.08\pm4.82$} & $0.897\pm0.092$ & \multicolumn{1}{c}{$32.25\pm4.74$} & $0.891\pm0.09$ \\ 
\bottomrule
\end{tabular}
}
\end{table}

\section{Conclusions}
We developed NPB-REC, a non-parametric Bayesian method for the reconstruction of brain MRI images from under-sampled k-space data with uncertainty estimation.
Specifically, we used noise injection for the training loss gradients to efficiently sample the true posterior distribution of the network weights. We used the E2E-VarNet network as a baseline. However, the proposed technique is not limited to a specific architecture and can be incorporated to any existing network.   
The conducted experiments showed that our approach enables uncertainty quantification that exhibits higher correlation with the reconstruction error than that obtained by Monte Carlo Dropout. In addition, it shows a significantly better reconstruction quality over other methods, especially with acceleration rate higher than that used in training. This demonstrates its ability to improve the generalization of the reconstruction over the other methods. 
\subsubsection{Acknowledgements} 
Khawaled, S. is a fellow of the Ariane de Rothschild Women Doctoral Program. Freiman , M. is a Taub fellow
(supported by the Taub Family Foundation, The Technion’s program for leaders in Science and Technology).
%
%
%

\bibliographystyle{splncs04}
\bibliography{refs}

\begin{thebibliography}{10}
\providecommand{\url}[1]{\texttt{#1}}
\providecommand{\urlprefix}{URL }
\providecommand{\doi}[1]{https://doi.org/#1}

\bibitem{akccakaya2019scan}
Ak{\c{c}}akaya, M., Moeller, S., Weing{\"a}rtner, S., U{\u{g}}urbil, K.:
  Scan-specific robust artificial-neural-networks for k-space interpolation
  (raki) reconstruction: Database-free deep learning for fast imaging. Magnetic
  resonance in medicine  \textbf{81}(1),  439--453 (2019)

\bibitem{avci2021quantifying}
Avci, M.Y., Li, Z., Fan, Q., Huang, S., Bilgic, B., Tian, Q.: Quantifying the
  uncertainty of neural networks using monte carlo dropout for deep learning
  based quantitative mri. arXiv preprint arXiv:2112.01587  (2021)

\bibitem{candes2006compressive}
Cand{\`e}s, E.J., et~al.: Compressive sampling. In: Proceedings of the
  international congress of mathematicians. vol.~3, pp. 1433--1452. Citeseer
  (2006)

\bibitem{cheng2019bayesian}
Cheng, Z., Gadelha, M., Maji, S., Sheldon, D.: A bayesian perspective on the
  deep image prior. In: Proceedings of the IEEE Conference on Computer Vision
  and Pattern Recognition. pp. 5443--5451 (2019)

\bibitem{edupuganti2020uncertainty}
Edupuganti, V., Mardani, M., Vasanawala, S., Pauly, J.: Uncertainty
  quantification in deep mri reconstruction. IEEE Transactions on Medical
  Imaging  \textbf{40}(1),  239--250 (2020)

\bibitem{eo2018kiki}
Eo, T., Jun, Y., Kim, T., Jang, J., Lee, H.J., Hwang, D.: Kiki-net:
  cross-domain convolutional neural networks for reconstructing undersampled
  magnetic resonance images. Magnetic resonance in medicine  \textbf{80}(5),
  2188--2201 (2018)

\bibitem{griswold2002generalized}
Griswold, M.A., Jakob, P.M., Heidemann, R.M., Nittka, M., Jellus, V., Wang, J.,
  Kiefer, B., Haase, A.: Generalized autocalibrating partially parallel
  acquisitions (grappa). Magnetic Resonance in Medicine: An Official Journal of
  the International Society for Magnetic Resonance in Medicine  \textbf{47}(6),
   1202--1210 (2002)

\bibitem{hammernik2018learning}
Hammernik, K., Klatzer, T., Kobler, E., Recht, M.P., Sodickson, D.K., Pock, T.,
  Knoll, F.: Learning a variational network for reconstruction of accelerated
  mri data. Magnetic resonance in medicine  \textbf{79}(6),  3055--3071 (2018)

\bibitem{khawaled2020unsupervised}
Khawaled, S., Freiman, M.: Unsupervised deep-learning based deformable image
  registration: a bayesian framework. arXiv preprint arXiv:2008.03949  (2020)

\bibitem{khawaled2022npbdreg}
Khawaled, S., Freiman, M.: Npbdreg: Uncertainty assessment in diffeomorphic
  brain mri registration using a non-parametric bayesian deep-learning based
  approach. Computerized Medical Imaging and Graphics p. 102087 (2022)

\bibitem{lustig2007sparse}
Lustig, M., Donoho, D., Pauly, J.M.: Sparse mri: The application of compressed
  sensing for rapid mr imaging. Magnetic Resonance in Medicine: An Official
  Journal of the International Society for Magnetic Resonance in Medicine
  \textbf{58}(6),  1182--1195 (2007)

\bibitem{majumdar_2015}
Majumdar, A.: Multi-Coil Parallel MRI Reconstruction, p. 86–119. Cambridge
  University Press (2015). \doi{10.1017/CBO9781316217795.005}

\bibitem{morris2018magnetic}
Morris, S.A., Slesnick, T.C.: Magnetic resonance imaging. Visual Guide to
  Neonatal Cardiology pp. 104--108 (2018)

\bibitem{pruessmann1999sense}
Pruessmann, K.P., Weiger, M., Scheidegger, M.B., Boesiger, P.: Sense:
  sensitivity encoding for fast mri. Magnetic Resonance in Medicine: An
  Official Journal of the International Society for Magnetic Resonance in
  Medicine  \textbf{42}(5),  952--962 (1999)

\bibitem{ronneberger2015u}
Ronneberger, O., Fischer, P., Brox, T.: U-net: Convolutional networks for
  biomedical image segmentation. In: International Conference on Medical image
  computing and computer-assisted intervention. pp. 234--241. Springer (2015)

\bibitem{shaul2020subsampled}
Shaul, R., David, I., Shitrit, O., Raviv, T.R.: Subsampled brain mri
  reconstruction by generative adversarial neural networks. Medical Image
  Analysis  \textbf{65},  101747 (2020)

\bibitem{sriram2020end}
Sriram, A., Zbontar, J., Murrell, T., Defazio, A., Zitnick, C.L., Yakubova, N.,
  Knoll, F., Johnson, P.: End-to-end variational networks for accelerated mri
  reconstruction. In: International Conference on Medical Image Computing and
  Computer-Assisted Intervention. pp. 64--73. Springer (2020)

\bibitem{tezcan2018mr}
Tezcan, K.C., Baumgartner, C.F., Luechinger, R., Pruessmann, K.P., Konukoglu,
  E.: Mr image reconstruction using deep density priors. IEEE transactions on
  medical imaging  \textbf{38}(7),  1633--1642 (2018)

\bibitem{welling2011bayesian}
Welling, M., Teh, Y.W.: Bayesian learning via stochastic gradient langevin
  dynamics. In: Proceedings of the 28th international conference on machine
  learning (ICML-11). pp. 681--688 (2011)

\bibitem{fastmri}
Zbontar, J., Knoll, F., Sriram, A., Murrell, T., Huang, Z., Muckley, M.J.,
  Defazio, A., Stern, R., Johnson, P., Bruno, M., et~al.: fastmri: An open
  dataset and benchmarks for accelerated mri. arXiv preprint arXiv:1811.08839
  (2018)

\bibitem{zhao2021fastmri}
Zhao, R., Yaman, B., Zhang, Y., Stewart, R., Dixon, A., Knoll, F., Huang, Z.,
  Lui, Y.W., Hansen, M., Lungren, M.P.: fastmri+: Clinical pathology
  annotations for knee and brain fully sampled multi-coil mri data. arXiv:
  Computer Vision and Pattern Recognition  \textbf{arXiv:2109.03812} (September
  2021)

\end{thebibliography}

\end{document}